# FABRICATION OF MINIATURIZED VARIABLE-FOCUS LENS USING LIQUID FILLING TECHNIQUE

*Hsiharng Yang[1], Chung-Yao Yang[1], Mau-Shiun Yeh[2]*

[1]Institute of Precision Engineering, National Chung Hsing University, Taichung, Taiwan 40227
[2]Materials & Electro-Optics Research Division, Chung-Shan Institute of Science & Technology, Tao-Yuan, Taiwan 325

**ABSTRACT**

This paper describes a simple method for fabricating a variable-focus lens by using PDMS (polydimethylsiloxane) and filling with liquid for the variable-focus lens. The lens diameter of 2-mm was designed in this experiment and expected to reach the focal length in the range of 3 ~ 12 mm. The theoretical value between the liquid volume and the lens contact angle at different focal lengths were simulated and measured. The pumped-in volumes ranged from 200 to 1400 μl, the contact angles ranged from 14.25° to 49.02°. Changing the deformation of PDMS film using different micro-fluidic volume produces the variable focal length from 4 ~ 10 mm in this experiment. The proposed method successfully fabricated a variable-focus lens. Bonding PDMS only once using no expensive instrument such as oxygen plasma was accomplished. The final objective is to insert the variable focus lens into portable optical imagery products.

## 1. INTRODUCTION

As imagery CCDs and CMOS detecting devices productions has gradually increased in recent years, the pixels of the detecting device and fabrication process yield rate have also improved. That resulted in promoting the development of the whole digital image storage industry. The primary lens today is the more complicated zoom lens made up of more than ten slices of optical glass. A single lens slice is called a lens element. Several lens elements make up an entire lens. Presently, the zoom lens is made primarily of a spherical convex lens, non-spherical convex lens, plane mirror or concave lens using different lens collocations. The focus of an entire lens is changed by the movement of the relative position between the lens elements. However, some zoom lenses are very large, heavy and difficult to carry. Present developments have tended towards thinner and lighter lens elements. The key components must be miniaturized to attain this goal. The single and independent variable focus lens has become the current fashion. Variable focus lens research has been carried out for a while.

Present research on variable focus lens can be divided into two driving methods. The first involves changing the lens radius of curvature using pressure. The second method involves varying with the applied voltage. In the pressure method, the PDMS membrane is the main parts of the variable focus lens. Chronis *et al.* fabricated the tunable liquid-filled microlens array integrated on top of a microfluidic network using soft lithography techniques [1]. The focal length of all microlenses is controlled using an elastomeric array by regulating the pressure in the micro-fluidics network. The structure is produced by jointing PDMS to glass using oxygen plasma. However, the jointing process is complicated and the glass is easily broken. Jeong *et al.* fabricated a tunable microdoublet lens array to form biconvex or meniscus lens [2]. It consisted of a tunable liquid-filled lens and a solid negative lens. The shortcoming is that the jointing process required expensive instruments like oxygen plasma. Chen *et al.* fabricated the tunable microlens consisting of a thin film with a 3D convex lens, chamber and microchannel made of PDMS [3]. The shortcomings were the jointing process being complicated and the alignment between the convex lens and microchannel. Agarwal *et al.* fabricated multi-level structures using lithography and ICP etching. The chamber was formed after jointing the multi-level structures [4]. Pressure is used to place tunable lens after lay over PDMS thin film on the chamber top and bottom. Pressure from the liquid filling in the microchannel is used to obtain variable focus in the optical lens system [5, 6].

Other variable-focus lens driven by the changing voltage can be divided into two methods. The first is electrowetting [7]. The second is liquid crystal rotation [8-12]. Kuipe used two immiscible liquids to form a meniscus. The first liquid is electrically conducting and the second insulates [7]. When a voltage is input, it will cause the contact angle of the interface between the two liquids to change, causing the focal length of the meniscus to vary. This method requires a high voltage input that may cause the liquids to vaporize. Ren *et al.* researched into variable-focus lens using liquid crystal molecules [8]. The liquid crystal will rotate when voltage is applied causing the refractive index to change. The necessary material is not easily to obtain and the structure is more complicated. This method is still research in progress [8-12].

This paper describes a simple method for fabricating a variable-focus lens using PDMS as the principal part for the variable-focus lens. PDMS has the advantages of elasticity and highly reproductivity. A simple fabrication process and common lithography instruments are proposed for producing such a variable-focus lens.

## 2. VARIABLE-FOCUS LENS DESIGN

A variable-focus lens with a diameter of 2-mm was designed in this experiment and expected to reach the focal length in the range of 3 ~ 12 mm. The theoretical value between the liquid volume and lens contact angle at different focal lengths was calculated using the lens formula. From thin lens equation [13],

$$\frac{1}{f} = (n-1)\left(\frac{1}{R_1} - \frac{1}{R_2}\right) \quad (1)$$

When $R_2 \to \infty$, the following equation is obtained.

$$R_1 = f(n-1) \quad (2)$$

A schematic diagram to sketch the contact angle related to the lens profile is shown in Figure 1. All symbols are listed in Table 1.

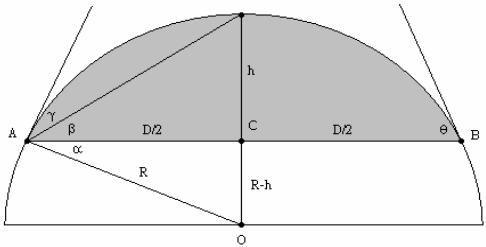

Figure 1. Schematic diagram showing contact angle and lens sag height in hemi-sphere.

The contact angle equation is calculated as follows:

$$R = \frac{D^2 + 4h^2}{8h} \quad (3)$$

$$\alpha = \tan^{-1}\left(\frac{R-h}{\frac{D}{2}}\right) \quad (4)$$

$$\beta = \tan^{-1}\left(\frac{h}{\frac{D}{2}}\right) \quad (5)$$

Then take Eq. (4) and (5) into the following equation

$$\gamma = 90° - \alpha - \beta = 90° - \tan^{-1}\left(\frac{2R-2h}{D}\right) - \tan^{-1}\left(\frac{2h}{D}\right), \quad (6)$$

and contact angle θ can be expressed as follows:

$$\theta = \gamma + \beta = 90° - \tan^{-1}\frac{2R-2h}{D}$$

$$= 90° - \tan^{-1}\left[\frac{2\left(\frac{D^2+4h^2}{8h}\right)-2h}{D}\right]$$

$$= 90° - \tan^{-1}\left[\frac{1-\left(\frac{h}{D}\right)^2}{4\frac{h}{D}}\right]. \quad (7)$$

From the following equation [14] the theoretical diagrams for V-f and V-θ as are obtained shown in Figure 2 and 3,

$$V = \frac{\pi R^3}{3}(2 + \cos\theta)(1 - \cos\theta)^2 \quad (8)$$

Table 1 List of symbols.

| Symbol | Description |
|---|---|
| n | Refractive index of medium |
| f | Focal length |
| R | Radius of curvature |
| D | Lens diameter |
| h | Lens sag height |
| θ | Contact angle |
| V | Pumped-in volume |

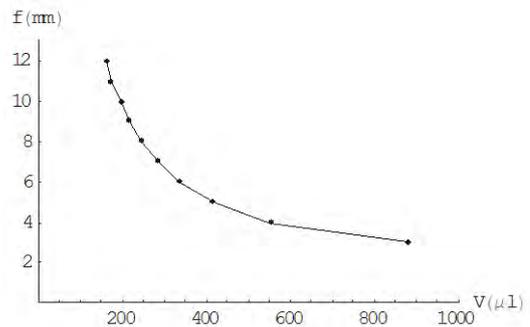

Figure 2. Relationship between the input volume and focal length (3 ~ 12 mm).

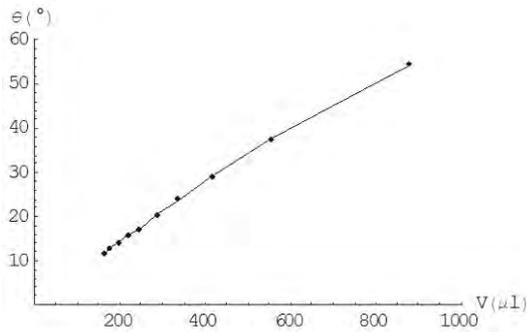

Figure 3. Relationship between the input volume and contact angle in different focal length (3 ~ 12 mm).

## 3. EXPERIMENTAL PROCESS

This study utilized PDMS elastomer as the main part of the variable-focus lens. The first mask is patterned with round and rectangular cavities (Figure 4) for the lens and fluidic channel. The second mask was designed to fabricate the cavity that stored the liquid and as the barricade to precisely control PDMS membrane. The process are shown in Figure 4 and described as follow.

First, JSR (THB-120N) negative acting photoresist is spin-coated onto a PMMA substrate at two-step speeds for about ten seconds at 150 rpm and 400 rpm, respectively. The device is then soft baked at a temperature of about 100 $^{o}$C for 6 minutes to dry the photoresist. The photoresist is then patterned through the mask and exposed to ultraviolet light using the conventional photolithography technique. The exposure dose was about 2000 mJ. After development, the first level structure is accomplished.

The above steps are repeated to form the second level circular chamber on top of the first layer after alignment with the second mask. After development, the substrate with a photoresist structure is heated at a temperature of about 120 $^{o}$C for 5 minutes. The two-layer fabrication process was developed to decouple the lens from the microchannel and thus minimize the distortion in the lens. The PDMS prepolymer mixture is spin cast over the mold at 400 rpm for 20 seconds. The PDMS mixture is subsequently cured in oven for 20 minutes at 140 $^{o}$C. The cured PDMS is carefully peeled off from the mold. The PDMS was then poured into the barricade, a piece of PDMS substrate to produce a thickness of about 1 ~ 2 mm after baking at temperature 140 $^{o}$C for 20 minutes in an oven.

The PDMS curing agent is spun coating on the PDMS substrate at 3000 rpm for 30 seconds resulting in a thickness of approximately 3 microns. The final structure is completed after the PDMS structure irreversibly bonded onto the PDMS substrate at 90 $^{o}$C for 20 minutes in oven. After proper trimming, the principal part of this variable-focus lens is accomplished. After that, the liquid input volumes are controlled using a micro pump or micro actuator. To change the deformation of the PDMS film by different microfluidic volume can achieve the desired variable focal length (Figure 5).

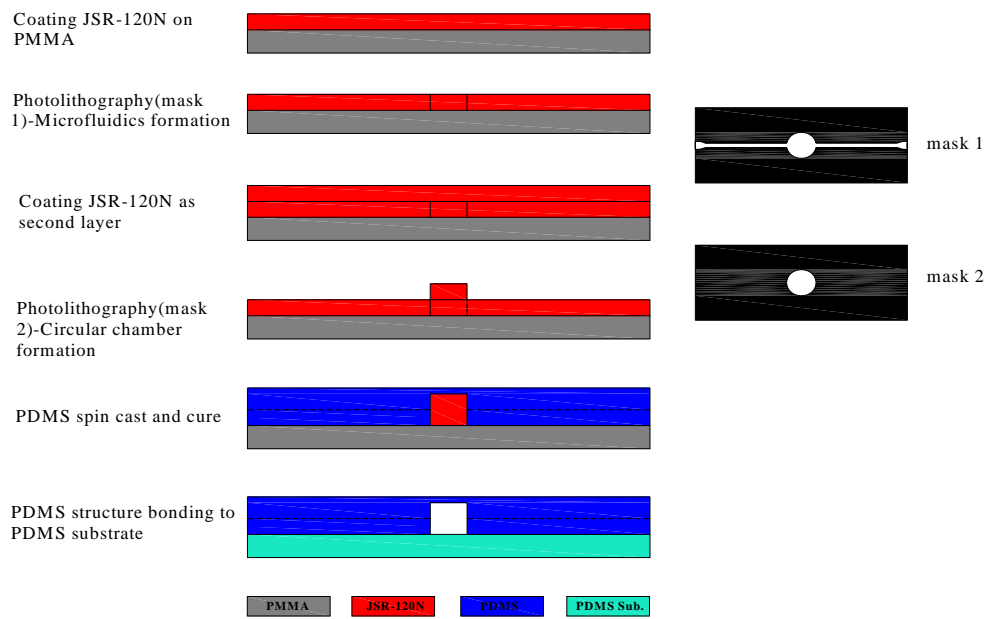

Figure 4. Main fabrication steps and the mask diagrams.

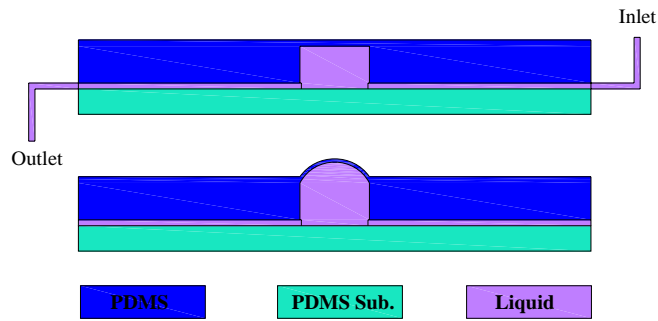

Figure 5. The acting diagram of the liquid filling variable-focus lens.

## 4. RESULTS AND DISCUSSION
### 4.1 Device fabrication

Figures 6(a) and 6(b) are SEM photographs of the lens structure in photoresist and the replication in PDMS structure, respectively. The 3D Confocal Microscopy was used to measure the height and 3D surface profile of the lens. The PDMS lens chamber is filled with deionized water with the refractive index 1.33. Each contact angle for a different pumped liquid volume is measured using a contact angle and surface tension instrument (FTA200). The pumped-in volume is controlled by a micro pump instrument in this experiment. Figure 9 shows the variant conditions of the contact angle for different pumped volumes. When the pumped-in volumes are 200, 400, 700, 1000, 1200 and 1400 μl, the contact angles are 14.25°, 22.63°, 28.56°, 38.07°, 43.22° and 49.02° in Figure 7(a) to Figure 7(f), respectively. And then take pumped-in volumes and contact angles into Equation (8) that can calculate the radius of curvature. Taking R into Equation (2) can get the focal length. The focal length ranged from 3.95 mm to 9.69 mm was calculated in this experiment. Figure 7(g) is the relationship of the pumped-in volume and the contact angle. The pump-in volume and the contact angle is approximately linear relationship as indicated in Figure 7(g). Figure 8 shows the relationship between the pumped-in volume and the focal length. Red dots in Figure 9 are the experimental data; then take curve fitting with the experimental data. Figure 9 shows the comparison between the theoretical and experimental values of the contact angle and the pumped volume. Figure 10 is the comparison between the theoretical and experimental values of the pumped volume and the focal length. The difference between two curves might be the loss of liquids due to possible leaking. The other reason may be air trapped in the chamber at the beginning, when the air is pressed out of the chamber after pumping. When liquids are filled up with the chamber, the pumped-in volume will lose a little. Therefore, the error may be improved by using some vacuum devices in the future.

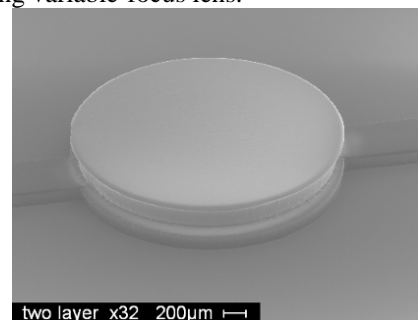

(a)

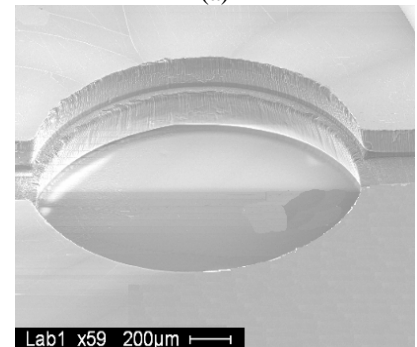

(b)

Figure 6. SEM photographs; (a) the photoresist structure, (b) PDMS structure.

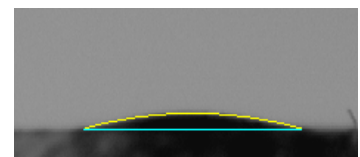

(a)

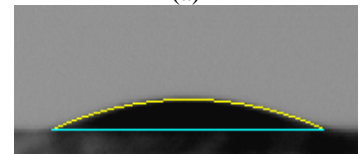

(b)

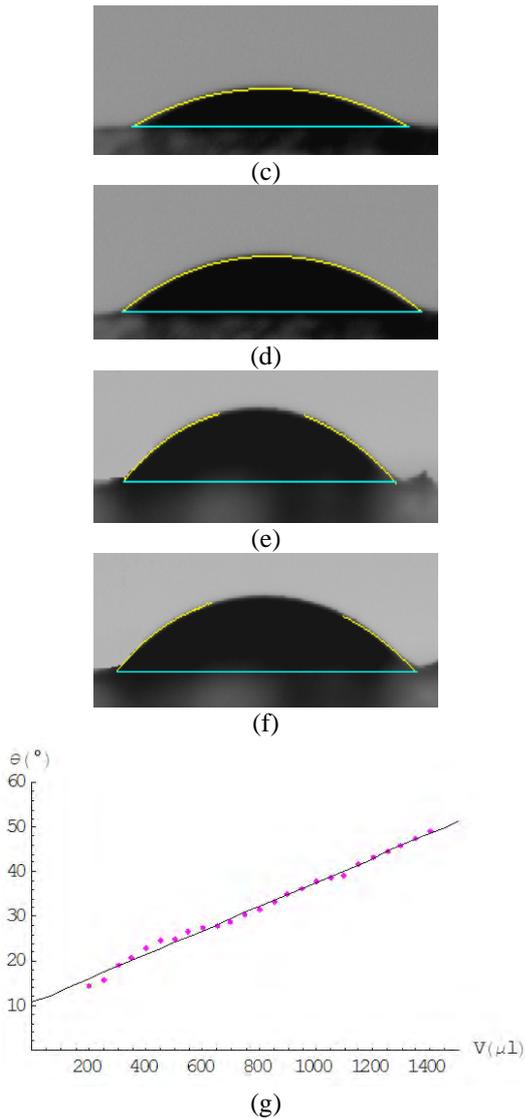

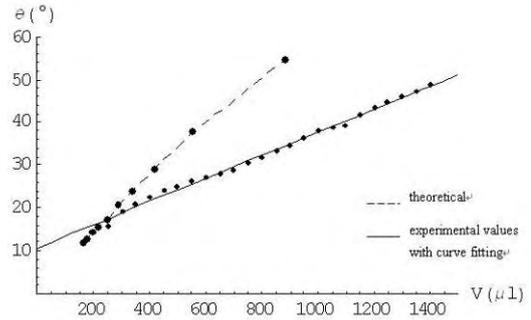

Figure 8. Relationship of the pumped-in volume and the focal length.

Figure 9. Comparison between theoretical and experimental values of the contact angle.

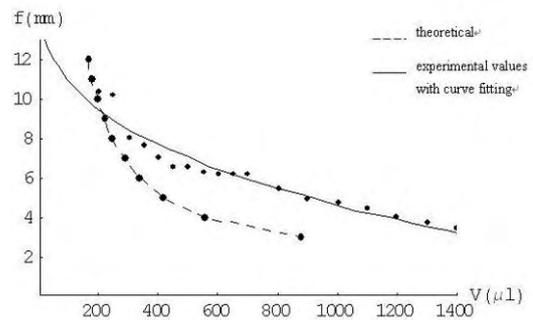

Figure 10. Comparison between theoretical and experimental values of the focal length.

**4.2 Optical performance**

Lens aberrations were simulated by using the commercial software, ZEMAX, which can build a model to analyze and assist in optical system design. Substitute the lens dimension and optical conditions to analyze focal length and circle of the least confusion (the envelope of convergent light can make up caustic, at the smallest radius of caustic is called circle of the least confusion) [13]. This area is regard as the best focus point or imaging point of the system with the aberration. The Image Space (F/#) is 2.8 which usually used in cell phone and the pumped-in liquid with refractive index 1.33 in this simulation. The simulated focal length is 10.83 mm when filling liquid at 200 μl. The simulated focal length is 3.43 mm when filling liquid at 1400 μl. Therefore, the variable-focus lens ranged 3.4 to 10.8 mm focal length was achieved in this simulation. The range is closed to that calculated by Equation (2) about 4 ~ 10 mm. The circles of the least confusion are 51.1 μm and 6.733 μm when pumped-in volumes at 200 μl and 1400 μl, respectively. From the results, the large pumped-in liquid generates the smaller circle of the least confusion. The smaller circle of the least confusion can provide the clearer imagery quality in the optical system. Figure 11 shows the variable focus lens in different pumped-in volumes. The magnified lens imaging photographs at different volumes were snapped by 3D optical microscopy. The high magnification was generated from the high pumped-in volume such as 1400 µl in Figure 11(c).

Figure 7. Photographs of the contact angle for different pumped-in volume measured using a contact angle and surface tension instrument (FTA200); (a) 200 μl, (b) 400 μl, (c) 700 μl, (d) 1000 μl, (e) 1200 μl, (f) 1400 μl, (g) relationship between the pumped-in volume and the contact angle.

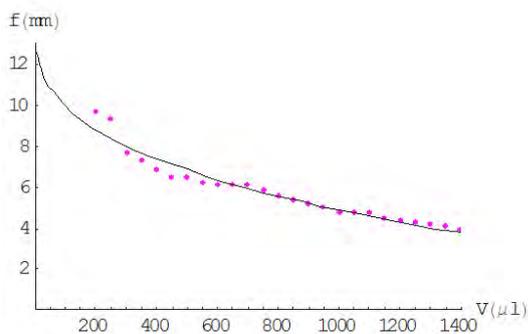

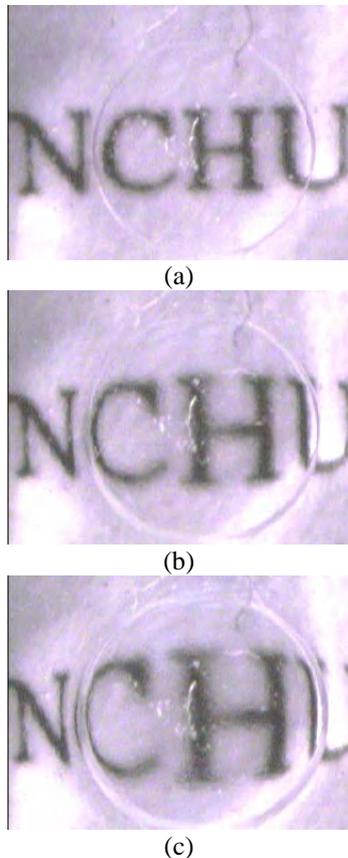

(a)

(b)

(c)

Figure 11. Imaging photographs of 2 mm lens at volume (a) 0µl, (b) 400µl, and (c) 1400.

## 5. CONCLUSION

This paper successfully demonstrated an easier method to fabricate a variable-focus lens. PDMS bonding was used only once and no expensive instruments such as oxygen plasma were used. The structure was fabricated with PDMS entirely to decrease the incompatibility between different materials. The structure does not easily burst because of its elasticity. The further objective is to insert the variable focus lens into portable optical imagery products. It also can be combined with CMOS or CCD for mobile phones, digital cameras, video cameras and medical microscopy. Each digital imaging storage media and cell phone communication can benefit from this technology. It also helps to advance micro optoelectromechanical system devices for the lower price trend in consuming electronic products.

## 6. ACKNOWLEDGEMENT

This work was supported by the National Science Council (series no. 94-2218-E-005-014) and Chung-Shan Institute of Science & Technology in Taiwan.